\documentclass[ALICE,manyauthors]{cernphprep}
\usepackage[comma,square,numbers,sort&compress]{natbib}
\usepackage{hyperref}
\usepackage{lineno}
\usepackage[T1]{fontenc}
\usepackage{xspace}
\usepackage[utf8]{inputenc}
\begin{document}
%%%%%%%%%%%%%%%%%%%%%%%%%%%%%%%%%%%%%%%%%%%%%%%%%%
% These are some new commands that may be useful 
% for paper writing in general. If other newcommands
% are needed for your specific paper, please feel 
% free to add here. 
%
% The currently available commands are organized in: 
% 1) Systems
% 2) Quantities
% 3) Energies and units
% 4) Detectors
% 5) particle species 
%%%%%%%%%%%%%%%%%%%%%%%%%%%%%%%%%%%%%%%%%%%%%%%%%%

\mathchardef\mhyphen="2D

% 1) SYSTEMS 
\newcommand{\pp}           {pp\xspace}
\newcommand{\ppbar}        {\mbox{$\mathrm {p\overline{p}}$}\xspace}
\newcommand{\XeXe}         {\mbox{Xe--Xe}\xspace}
\newcommand{\PbPb}         {\mbox{Pb--Pb}\xspace}
\newcommand{\pA}           {\mbox{pA}\xspace}
\newcommand{\pPb}          {\mbox{p--Pb}\xspace}
\newcommand{\AuAu}         {\mbox{Au--Au}\xspace}
\newcommand{\dAu}          {\mbox{d--Au}\xspace}

% 2) QUANTITIES 
\newcommand{\sigmapid}{$\sigma^{^{3}\mathrm{He}}_{\mathrm{d}E/\mathrm{d}x}$}
\newcommand{\nsigma} {$\left(  \mathrm{d}E/\mathrm{d}x -  \langle \mathrm{d}E/\mathrm{d}x \rangle_{^{3}\mathrm{He}} \right) /\sigma^{^{3}\mathrm{He}}_{\mathrm{d}E/\mathrm{d}x}$ }
\newcommand{\s}            {\ensuremath{\sqrt{s}}\xspace}
\newcommand{\snn}          {\ensuremath{\sqrt{s_{\mathrm{NN}}}}\xspace}
\newcommand{\pt}           {\ensuremath{p_{\rm T}}\xspace}
\newcommand{\meanpt}       {$\langle p_{\mathrm{T}}\rangle$\xspace}
\newcommand{\ycms}         {\ensuremath{y_{\rm CMS}}\xspace}
\newcommand{\ylab}         {\ensuremath{y_{\rm lab}}\xspace}
\newcommand{\etarange}[1]  {\mbox{$\left | \eta \right |~<~#1$}}
\newcommand{\yrange}[1]    {\mbox{$\left | y \right |~<~#1$}}
\newcommand{\dndy}         {\ensuremath{\mathrm{d}N_\mathrm{ch}/\mathrm{d}y}\xspace}
\newcommand{\dndeta}       {\ensuremath{\mathrm{d}N_\mathrm{ch}/\mathrm{d}\eta}\xspace}
\newcommand{\avdndeta}     {\ensuremath{\langle\dndeta\rangle}\xspace}
\newcommand{\dNdy}         {\ensuremath{\mathrm{d}N_\mathrm{ch}/\mathrm{d}y}\xspace}
\newcommand{\Npart}        {\ensuremath{N_\mathrm{part}}\xspace}
\newcommand{\Ncoll}        {\ensuremath{N_\mathrm{coll}}\xspace}
\newcommand{\dEdx}         {\ensuremath{\textrm{d}E/\textrm{d}x}\xspace}
\newcommand{\RpPb}         {\ensuremath{R_{\rm pPb}}\xspace}

% 3) ENERGIES, UNITS
\newcommand{\nineH}        {$\sqrt{s}~=~0.9$~Te\kern-.1emV\xspace}
\newcommand{\seven}        {$\sqrt{s}~=~7$~Te\kern-.1emV\xspace}
\newcommand{\twoH}         {$\sqrt{s}~=~0.2$~Te\kern-.1emV\xspace}
\newcommand{\twosevensix}  {$\sqrt{s}~=~2.76$~Te\kern-.1emV\xspace}
\newcommand{\five}         {$\sqrt{s}~=~5.02$~Te\kern-.1emV\xspace}
\newcommand{\twosevensixnn}{$\sqrt{s_{\mathrm{NN}}}~=~2.76$~Te\kern-.1emV\xspace}
\newcommand{\fivenn}       {$\sqrt{s_{\mathrm{NN}}}~=~5.02$~Te\kern-.1emV\xspace}
\newcommand{\LT}           {L{\'e}vy-Tsallis\xspace}
\newcommand{\GeVc}         {Ge\kern-.1emV/$c$\xspace}
\newcommand{\MeVc}         {Me\kern-.1emV/$c$\xspace}
\newcommand{\TeV}          {Te\kern-.1emV\xspace}
\newcommand{\GeV}          {Ge\kern-.1emV\xspace}
\newcommand{\MeV}          {Me\kern-.1emV\xspace}
\newcommand{\GeVmass}      {Ge\kern-.2emV/$c^2$\xspace}
\newcommand{\MeVmass}      {Me\kern-.2emV/$c^2$\xspace}
\newcommand{\lumi}         {\ensuremath{\mathcal{L}}\xspace}

% 4) DETECTORS 
\newcommand{\ITS}          {\rm{ITS}\xspace}
\newcommand{\TOF}          {\rm{TOF}\xspace}
\newcommand{\ZDC}          {\rm{ZDC}\xspace}
\newcommand{\ZDCs}         {\rm{ZDCs}\xspace}
\newcommand{\ZNA}          {\rm{ZNA}\xspace}
\newcommand{\ZNC}          {\rm{ZNC}\xspace}
\newcommand{\SPD}          {\rm{SPD}\xspace}
\newcommand{\SDD}          {\rm{SDD}\xspace}
\newcommand{\SSD}          {\rm{SSD}\xspace}
\newcommand{\TPC}          {\rm{TPC}\xspace}
\newcommand{\TRD}          {\rm{TRD}\xspace}
\newcommand{\VZERO}        {\rm{V0}\xspace}
\newcommand{\VZEROA}       {\rm{V0A}\xspace}
\newcommand{\VZEROC}       {\rm{V0C}\xspace}
\newcommand{\Vdecay} 	   {\ensuremath{V^{0}}\xspace}

% 4) PARTICLE SPECIES 
\newcommand{\ee}           {\ensuremath{e^{+}e^{-}}} 
\newcommand{\pip}          {\ensuremath{\pi^{+}}\xspace}
\newcommand{\pim}          {\ensuremath{\pi^{-}}\xspace}
\newcommand{\kap}          {\ensuremath{\rm{K}^{+}}\xspace}
\newcommand{\kam}          {\ensuremath{\rm{K}^{-}}\xspace}
\newcommand{\pbar}         {\ensuremath{\rm\overline{p}}\xspace}
\newcommand{\kzero}        {\ensuremath{{\rm K}^{0}_{\rm{S}}}\xspace}
\newcommand{\lmb}          {\ensuremath{\Lambda}\xspace}
\newcommand{\almb}         {\ensuremath{\overline{\Lambda}}\xspace}
\newcommand{\Om}           {\ensuremath{\Omega^-}\xspace}
\newcommand{\Mo}           {\ensuremath{\overline{\Omega}^+}\xspace}
\newcommand{\X}            {\ensuremath{\Xi^-}\xspace}
\newcommand{\Ix}           {\ensuremath{\overline{\Xi}^+}\xspace}
\newcommand{\Xis}          {\ensuremath{\Xi^{\pm}}\xspace}
\newcommand{\Oms}          {\ensuremath{\Omega^{\pm}}\xspace}
\newcommand{\degree}       {\ensuremath{^{\rm o}}\xspace}

%%%%%%%%%%%%%%%  Title page %%%%%%%%%%%%https://www.overleaf.com/1626559919pmsqkkvdjyzj%%%%%%%%%%%%%%%%%%%
\begin{titlepage}
% the dates below correspond to CERN approval
% please don't touch: EB chairs will take care CERN-EP-2019-202
\PHyear{2019}       % required, will be obtained from CERN
\PHnumber{202}      % required, will be obtained from CERN
\PHdate{18 September}  % required, will be obtained from CERN
%%%%%%%%%%%%%%%%%%%%%%%%%%%%%%%%%%%%%%%%%%%%%%%%%%%%

%%% Put your own title + short title here:
\title{Measurement of the (anti-)$^{3}$He elliptic flow in \PbPb collisions \newline at \snn = 5.02 TeV}
\ShortTitle{$^{3}$He elliptic flow in \PbPb collisions}   % appears on left page headers

%%% Do not change the next lines
\Collaboration{ALICE Collaboration\thanks{See Appendix~\ref{app:collab} for the list of collaboration members}}
\ShortAuthor{ALICE Collaboration} % appears on right page headers, do not change

\begin{abstract}

The elliptic flow ($v_{2}$) of (anti-)$^{3}$He is measured in \PbPb collisions at \snn = 5.02 TeV in the transverse-momentum (\pt) range of 2--6 GeV/$\textit{c}$ for the centrality classes 0--20$\%$, \mbox{20--40$\%$}, and \mbox{40--60$\%$} using the event-plane method. 
This measurement is compared to that of pions, kaons, and protons at the same center-of-mass energy. A clear mass ordering is observed at low \pt, as expected from relativistic hydrodynamics. 
The violation of the scaling of $v_{2}$ with the number of constituent quarks at low \pt, already observed for identified hadrons and deuterons at LHC energies, is confirmed also for (anti-)$^{3}$He. 
The elliptic flow of (anti-)$^{3}$He is underestimated by the Blast-Wave model and overestimated by a simple coalescence approach based on nucleon scaling. 
The elliptic flow of (anti-)$^{3}$He measured in the centrality classes 0--20$\%$ and 20--40$\%$ is well described by a more sophisticated coalescence model where the phase-space distributions of protons and neutrons are generated using the iEBE-VISHNU hybrid model with AMPT initial conditions.

\end{abstract}
\end{titlepage}

\setcounter{page}{2}

\section{Introduction} 
\label{sec:Introduction}

The primary goal of studying ultra-relativistic heavy-ion collisions is to investigate the properties of the Quark--Gluon Plasma (QGP), a phase of matter made of deconfined quarks and gluons, which is created under extreme conditions of high temperature and energy density. At the Large Hadron Collider (LHC), the QGP can be studied in a region of the phase diagram where a cross-over transition from the deconfined phase to ordinary nuclear matter is expected based on Quantum Chromodynamics (QCD) calculations on the lattice \cite{LatticeQCD1,LatticeQCD2,LatticeQCD3}. 

In ultra-relativistic heavy-ion collisions, light nuclei, hypernuclei, and their antiparticles are produced in addition to other particle species. 
The production mechanism of these loosely bound composite objects in heavy-ion collisions is not clear and is still under debate.
Two phenomenological models are typically used to describe the light (anti-)(hyper-)nuclei production: the statistical hadronization model \cite{SHM1,SHM2,SHM3,SHM4,SHM5,SHM6} and the coalescence approach \cite{Coalescence1,Coalescence2,Coalescence3,iEBE_VISHNU}. In the former, light nuclei are assumed to be emitted by a source in local thermal and hadrochemical equilibrium and their abundances are fixed at chemical freeze-out. This model reproduces the light-flavored hadron yields measured in central nucleus--nucleus collisions, including those of (anti-)nuclei and (anti-)hypernuclei \cite{SHM1}. 
However, the detailed mechanism of hadron production and the explanation of the propagation of loosely-bound states through the hadron gas phase without a significant reduction in their yields are not addressed by this model. It has been conjectured that such objects could be produced at the phase transition as compact colorless quark clusters which are expected to interact little with the surrounding matter \cite{SHM5}. 
In the coalescence approach, light nuclei are assumed to be formed by the coalescence of protons and neutrons which are close in phase-space at kinetic freeze-out \cite{Coalescence2}. In the simple version of this model, nucleons are treated as point-like particles and the coalescence process is assumed to happen if the difference between their momenta is smaller than a given threshold, typically of the order of 100 MeV/$\textit{c}$, which is a free parameter of the model, while space coordinates are ignored. 
On the contrary, in the state-of-the-art implementations of the coalescence approach, the quantum-mechanical properties of nucleons and nuclei are taken into account and the coalescence probability is calculated from the overlap between the wave functions of protons and neutrons which are mapped onto the Wigner density of the nucleus.
The phase-space distributions of protons and neutrons at the kinetic freeze-out are generated from particle production models, such as A Multi-Phase Transport Model (AMPT) \cite{AMPT}, or from hydrodynamical simulations coupled to hadronic transport models \cite{iEBE_VISHNU}.  
The advanced coalescence model qualitatively describes the deuteron-to-proton and $^{3}$He-to-proton ratios measured in different collision systems as a function of the charged-particle multiplicity \cite{coalescenceSmallSystems}, while the simple coalescence approach provides a description of \pt spectra of light (anti-)nuclei measured in high-energy hadronic collisions only in the low-multiplicity regime \cite{deuteron_pPbALICE}. 

A key observable to study the production mechanism of light (anti-)nuclei is the elliptic flow, i.e. the second harmonic ($v_{2}$) of the Fourier decomposition of their azimuthal production distribution with respect to a collision symmetry plane. The latter is defined by the impact parameter of the incoming nuclei and the beam direction \cite{AnisotropyAndFlow}.
The elliptic flow of light nuclei was measured by PHENIX \cite{deuteronFlowPHENIX} and STAR \cite{lightNucleiFlowSTAR} at the Relativistic Heavy Ion Collider (RHIC). 
The centrality dependence of $v_{2}$ for deuterons (d) and antideuterons ($\overline{\mathrm{d}}$) was found to be qualitatively similar to that of identified hadrons \cite{lightNucleiFlowSTAR}. 
An approximate atomic mass number ($A$) scaling was observed for the elliptic flow of light nuclei when compared to the proton $v_{2}$ up to $p_{\mathrm{T}}/A = 1.5\ \mathrm{GeV}/\textit{c}$, with slight deviations for higher $p_{\mathrm{T}}/A$ \cite{lightNucleiFlowSTAR}. 
The flow of identified hadrons is often described using the Blast-Wave model \cite{BlastWave1,BlastWave2,BlastWave3}. This is a model inspired by hydrodynamics, which assumes that the system produced in heavy-ion collisions is locally thermalized and expands collectively with a common velocity field. The system undergoes a kinetic freeze-out at the temperature $T_{\mathrm{kin}}$ and is characterized by a common transverse radial flow velocity ($\beta$) at the freeze-out surface.
The Blast-Wave model, however, fails in reproducing the $v_{2}$ of light nuclei measured in \AuAu collisions at \snn = 200 GeV \cite{lightNucleiFlowSTAR}, which is instead well described by a more sophisticated coalescence model where the phase-space distributions of nucleons are generated using the string-melting version of AMPT \cite{AMPT}. 

The elliptic flow of d and $\overline{\mathrm{d}}$ was measured by the ALICE Collaboration in \PbPb collisions at \snn = 2.76 TeV in the transverse-momentum range \mbox{$0.8\ \leq p_{\mathrm{T}} < 5\ \mathrm{GeV}/\textit{c}$} for different centrality classes \cite{deuteronFlowALICE}. The scaling of $v_{2}$ with the number of constituent quarks ($n_{\mathrm{q}}$) is violated for identified hadrons including deuterons, with deviations up to $20\%$ \cite{deuteronFlowALICE}.
Predictions from simultaneous fits of the \pt spectra and the $v_{2}$ of charged pions, kaons, and protons using a Blast-Wave model provide a good description of the $v_{2}$ of deuterons in the measured \pt range for all centralities, consistent with common kinetic freeze-out conditions \cite{deuteronFlowALICE}. A simple coalescence model, based on the $A$-scaling of $v_{2}$ \cite{SimpleCoalescenceFlow}, fails in reproducing the data for all centralities and in the entire \pt range \cite{deuteronFlowALICE}. 
The data are fairly well described by a coalescence approach which uses as an input the phase-space distributions generated with the default AMPT settings \cite{iEBE_VISHNU}. However, this model does not describe the coalescence parameter $B_{2}$, defined as the ratio between the invariant yield of deuterons and the square of the invariant yield of protons \cite{deuteronFlowALICE}. The predictions obtained using the string-melting version of AMPT, which described RHIC data, are not consistent with the ALICE measurement \cite{deuteronFlowALICE}. 

The first measurement of the (anti-)$^{3}$He elliptic flow in \PbPb collisions at \snn = 5.02 TeV is presented in this paper. This measurement complements the picture obtained from that of the proton and deuteron flow at LHC energies. 
\section{Experimental apparatus and data sample} 
\label{sec:ExperimentalApparatus}

ALICE is one of the four big experiments at the LHC dedicated to the study of heavy-ion collisions at ultra-relativistic energies. A detailed description of the ALICE apparatus and its performance can be found in Refs. \cite{ALICEexperiment} and \cite{ALICEperformance}. 

Trajectories of charged particles are reconstructed in the ALICE central barrel with the Inner Tracking System (ITS) \cite{ALICEexperiment} and the Time Projection Chamber (TPC) \cite{TPC}. These are located within a large solenoidal magnet, providing a highly homogeneous magnetic field of 0.5 T parallel to the beam line. 
The ITS consists of six cylindrical layers of silicon detectors with a total pseudorapidity coverage $|\eta| < 0.9$ with respect to the nominal interaction region. The ITS is used in the determination of primary and secondary vertices, and in the track reconstruction.
The TPC is the largest detector in the ALICE central barrel, with a pseudorapidity coverage $|\eta| < 0.9$. It is used for track reconstruction, charged-particle momentum measurement and for particle identification via the measurement of the specific energy loss of particles in the TPC gas. 
The transverse-momentum resolution ranges from about 1$\%$ at 1 GeV/$\textit{c}$ to about 10$\%$ at 50 GeV/$\textit{c}$ in \PbPb collisions at \snn = 2.76 TeV \cite{ALICEperformance} and at \snn = 5.02 TeV \cite{ptSpectraRAA}. The \dEdx resolution depends on centrality and is in the range 5--6.5$\%$ for minimum ionizing particles crossing the full volume of the TPC \cite{ALICEperformance}. 
Collision events are triggered by two plastic scintillator arrays, V0A and V0C \cite{VZEROPerformance}, located on both sides of the interaction point, covering the pseudorapidity regions $-3.7 <  \eta < -1.7$  and \mbox{2.8 $< \eta <$ 5.1}. Each V0 array consists of four rings in the radial direction, with each ring comprising eight cells with the same azimuthal size. The V0 scintillators are used to determine the collision centrality from the measured charged-particle multiplicity \cite{CentralityDetermination,CentralityDeterminationPublicNote}, and to measure the orientation of the symmetry plane of the collision. 

The data used for this analysis were collected in 2015 during the LHC \PbPb run at \snn = 5.02 TeV. A minimum bias event trigger was used, which requires coincident signals in the V0 detectors synchronous with the bunch crossing time defined by the LHC clock.

\section{Data analysis} 
\label{sec:DataAnalysis}

\subsection{Event selection} 
\label{subsec:EventSelection}

In order to keep the conditions of the detectors as uniform as possible and reject background collisions, the coordinate of the primary vertex along the beam axis is required to be within 10 cm from the nominal interaction point. Collisions with multiple primary vertices are tagged as pile-up events and rejected.
A centrality-dependent non-uniformity in the angular distribution of the symmetry plane, of maximum $6\%$ is found. In order to correct for this non-uniformity, the events are re-weighted based on the collision centrality ($C$) and the angle of the symmetry plane $\Psi_{\mathrm{2}}$. 
The weight for a given two-dimensional cell ($C$,$\Psi_{\mathrm{2}}$) is defined as the ratio between the average number of events, for all $C$ and $\Psi_{\mathrm{2}}$, and the actual number of events in the same two-dimensional cell.
The centrality classes used for the analysis presented in this Letter are 0--20$\%$, 20--40$\%$, and 40--60$\%$.
In total, approximately 20 million events are selected in each centrality class.

\subsection{Track selection and particle identification} 
\label{subsec:TrackSelectionAndParticleIdentification}

(Anti-)$^{3}$He candidates are selected from the charged-particle tracks reconstructed in the ITS and TPC in the kinematic range \pt$/|z|$ $> 1\ \mathrm{GeV}/\textit{c}$ and $|\eta| < 0.8$, where $z$ is the particle electric charge in units of the elementary charge. 
Tracks are required to have a minimum number of clusters in the TPC, $N_{\mathrm{cls}}^{\mathrm{TPC}}$, of at least 70 out of a maximum of 159, and in the ITS, $N_{\mathrm{cls}}^{\mathrm{ITS}}$, of at least two with one cluster located in any of the two innermost ITS layers. The number of TPC clusters used in the \dEdx calculation, $N_{\mathrm{cls}}^{\mathrm{TPC}} \left( \mathrm{d}E/\mathrm{d}x \right)$ is required to be larger than 50.
Good quality of the track fit is also required, expressed by $\chi^{2}/N_{\mathrm{cls}}^{\mathrm{TPC}}<4$ and a ratio of the number of TPC clusters attached to the track over the number of findable TPC clusters (accounting for track length, location, and momentum) larger than 80$\%$. 
The contribution from secondary tracks is reduced by requiring a maximum Distance of Closest Approach (DCA) to the primary vertex in the transverse plane ($\mathrm{DCA}_{\mathrm{xy}}<0.1$ cm) and in the longitudinal direction ($\mathrm{DCA}_{\mathrm{z}}<1$ cm).
These selection criteria ensure a high track-reconstruction efficiency, which is larger than $80\%$, and a resolution in the \dEdx measured in the TPC of about $6\%$ in the centrality and \pt ranges used for this measurement. 

\begin{figure}[hbt]
    \begin{center}
    \includegraphics[width = 0.9\textwidth]{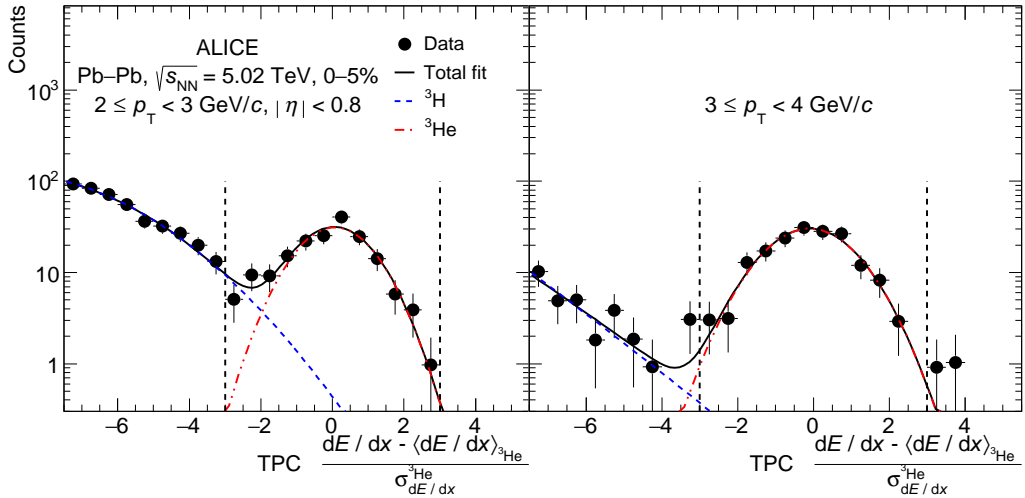}
    \end{center}
    \caption{Distributions of \nsigma measured in the TPC for the transverse-momentum ranges $2\ \leq p_{\mathrm{T}} < 3\ \mathrm{GeV}/\textit{c}$ (left) and $3\ \leq p_{\mathrm{T}} < 4\ \mathrm{GeV}/\textit{c}$ (right). The vertical bars represent the statistical uncertainties of the data. The blue dotted and the red dash-dotted lines indicate the $^{3}$H and $^{3}$He contributions while the black solid lines show the sum of both. The ranges used for the signal extraction are indicated by the vertical black-dotted lines. }
    \label{fig:ParticleIdentification}
\end{figure}

The expected average \dEdx for (anti-)$^{3}$He, $\langle \mathrm{d}E/\mathrm{d}x \rangle_{^{3}\mathrm{He}}$, is given by the Bethe formula and the standard deviation of the distribution of \dEdx $-$ $\langle \mathrm{d}E/\mathrm{d}x \rangle_{^{3}\mathrm{He}}$, denoted \sigmapid, is the TPC \dEdx resolution measured for (anti-)$^{3}$He.  
For the (anti-)$^{3}$He identification, the \dEdx measured in the TPC is required to be within 3\sigmapid from the expected average for $^{3}$He.
The distributions of \nsigma for the transverse-momentum ranges $2\ \leq p_{\mathrm{T}} < 3\ \mathrm{GeV}/\textit{c}$ and $3\ \leq p_{\mathrm{T}} < 4\ \mathrm{GeV}/\textit{c}$ are shown in Fig.~\ref{fig:ParticleIdentification}. The range used for the (anti-)$^{3}$He selection is indicated by the vertical black-dotted lines.
The contamination by (anti-)$^{3}$H is estimated by fitting the measured \nsigma distribution in a given \pt range using two Gaussian functions, one for (anti-)$^{3}$H and the other for (anti-)$^{3}$He. The (anti-)$^{3}$H contribution is subtracted from the distribution to extract the (anti-)$^{3}$He signal in the range within $\pm$3\sigmapid. The contamination from (anti-)$^{3}$H is negligible for $p_{\mathrm{T}} > 3\ \mathrm{GeV}/\textit{c}$ (see right panel of Fig.~\ref{fig:ParticleIdentification}). The contamination from (anti-)$^{4}$He is expected to be negligible over the full \pt range considering that its production rate measured in \PbPb collisions at \snn = 2.76 TeV is suppressed compared to that of (anti-)$^{3}$He by a factor $\sim 300$ \cite{Alpha}.

\subsection{Secondary \texorpdfstring{$^{3}$}{}He from spallation processes  } 
\label{subsec:SecondaryHeliumFromSpallation}

\begin{figure}[hbt]
    \begin{center}
    \includegraphics[width = 0.6\textwidth]{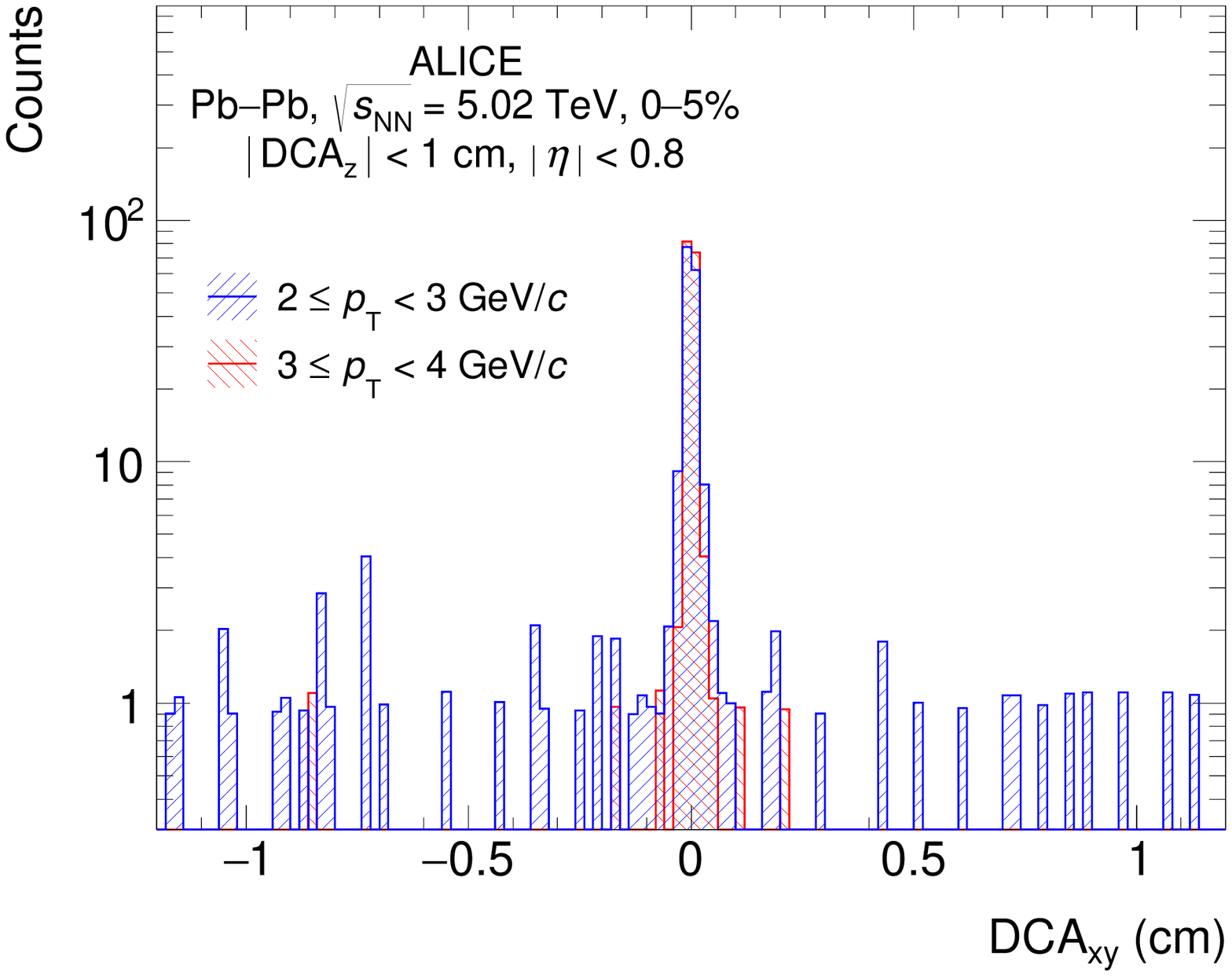}
    \end{center}
    \caption{$\mathrm{DCA}_{\mathrm{xy}}$ distributions of $^{3}$He candidates, selected requiring $-2$ $<$ \nsigma $<$ 3, with $|\mathrm{DCA}_{\mathrm{z}}| < 1$ cm measured in the transverse-momentum intervals $2\ \leq p_{\mathrm{T}} < 3\ \mathrm{GeV}/\textit{c}$ (blue) and $3\ \leq p_{\mathrm{T}} < 4\ \mathrm{GeV}/\textit{c}$ (red). }
    \label{fig:DCAdistributions}
\end{figure}

The main background for this measurement is represented by secondary $^{3}$He produced by spallation reactions in the interactions between primary particles and nuclei in the detector material or in the beam pipe. 
This background source is relevant only for $^{3}$He, while this effect is negligible for anti-$^{3}$He. Nuclear fragments emitted in spallation processes have almost uniform angular distributions with respect to the direction of the incoming particle, while primary $^{3}$He tracks originate from the primary vertex.
The contribution of secondary $^{3}$He produced by spallation can be investigated from the $\mathrm{DCA}_{\mathrm{xy}}$ distribution, which has a peak around zero for primary $^{3}$He and is almost flat for secondary $^{3}$He.    
The $\mathrm{DCA}_{\mathrm{xy}}$ distributions for $^{3}$He candidates measured in the transverse-momentum ranges $2\ \leq p_{\mathrm{T}} < 3\ \mathrm{GeV}/\textit{c}$ and $3\ \leq p_{\mathrm{T}} < 4\ \mathrm{GeV}/\textit{c}$ are shown in Fig.~\ref{fig:DCAdistributions}. The sign of the $\mathrm{DCA}_{\mathrm{xy}}$ is positive if the primary vertex is inside the track curvature and negative if it lies outside. These distributions are obtained by selecting tracks with $|\mathrm{DCA}_{\mathrm{z}}| <$ 1 cm and applying a stricter requirement for the selection of $^{3}$He candidates, given by $-2 \leq$ \nsigma $<3$. This asymmetric range is used to increase the purity of the $^{3}$He sample by suppressing the $^{3}$H contamination. The contribution from secondary $^{3}$He produced by spallation is found to be relevant in this analysis only in the transverse-momentum range $2\ \leq p_{\mathrm{T}} < 3\ \mathrm{GeV}/\textit{c}$.

For the measurement presented in this Letter, $^{3}\overline{\mathrm{He}}$ are used for $2\ \leq p_{\mathrm{T}} < 3\ \mathrm{GeV}/\textit{c}$, while the sum of $^{3}$He and $^{3}\overline{\mathrm{He}}$ is used for higher \pt where the contribution from secondary $^{3}$He from spallation is negligible. This is possible because the elliptic flow of $^{3}\mathrm{He}$ and $^{3}\overline{\mathrm{He}}$ are consistent within the statistical uncertainties in the \pt range where these two measurements can be compared, i.e. $p_{\mathrm{T}} > 3\ \mathrm{GeV}/\textit{c}$, and in all centrality intervals. A vanishing difference between the elliptic flow of matter and antimatter nuclei at LHC energies is already observed for (anti-)protons \cite{protonV2_2760,protonV2} and (anti-)deuterons \cite{deuteronFlowALICE}. This observation is consistent with the decreasing trend of the difference between the elliptic flow of protons and antiprotons, deuterons and antideuterons with increasing center-of-mass energy at RHIC going from \snn = 7.7 GeV to \snn = 200 GeV \cite{reviewLightNucleiFlow}.

\subsection{The event-plane method} 
\label{subsec:TheEventPlaneMethod}

The initial spatial anisotropy of the hot and dense matter created in non-central nucleus--nucleus collisions results in an azimuthal anisotropy of particle emission with respect to the symmetry plane. The azimuthal distribution of the emitted particles can be expressed as a Fourier series \cite{FourierDecomposition} 

\begin{equation}
\frac{\mathrm{d}N}{\mathrm{d}\varphi} \propto 1 + 2\sum_{n \geq 1} v_{n} \mathrm{cos} \big( n \big(\varphi - \Psi_{\mathrm{n}} \big) \big),
\end{equation}

where $\Psi_{\mathrm{n}}$ indicates the orientation of the $n$th symmetry plane, $\varphi$ is the azimuthal angle of a particle, and the Fourier coefficients $v_{n}$ are also referred to as the flow coefficients.

Experimentally, the true symmetry plane can only be reconstructed approximately because of the finite detector resolution. The measured symmetry plane is called `event plane'.
The elliptic flow of (anti-)$^{3}$He is measured using the Event-Plane (EP) method \cite{EventPlaneMethod}. The $v_{2}$ of (anti-)$^{3}$He in each \pt range is given by

\begin{equation}
v_{2} \{\mathrm{EP}, |\Delta \eta|>0.9\}  \left( p_{\mathrm{T}} \right) = \frac{\pi}{4R_{\Psi_{2}}} \frac{N_{\mathrm{in\mhyphen plane}} \left( p_{\mathrm{T}} \right) - N_{\mathrm{out\mhyphen of\mhyphen plane}} \left( p_{\mathrm{T}} \right)}{N_{\mathrm{in\mhyphen plane}} \left( p_{\mathrm{T}} \right) + N_{\mathrm{out\mhyphen of\mhyphen plane}} \left( p_{\mathrm{T}} \right)},
\label{Eq:EllipticFlow}
\end{equation}

where the $|\Delta \eta|$ represents the minimal pseudorapidity gap between the V0 detectors and the TPC, the $R_{\Psi_{2}}$ is the event-plane resolution for the second harmonic, and $N_{\mathrm{in\mhyphen plane}}$ and $N_{\mathrm{out\mhyphen of\mhyphen plane}}$ are the number of (anti-)$^{3}$He candidates in-plane and out-of-plane, respectively. Particles are regarded as `in-plane' if the azimuthal difference $|\Delta \varphi | = |  \varphi - \Psi_{2}^{\mathrm{EP}}|  < 45^{\circ} $ or $|\Delta \varphi | = |\varphi - \Psi_{2}^{\mathrm{EP}}|  > 135^{\circ}$, and `out-of-plane' otherwise, where $\Psi_{2}^{\mathrm{EP}}$ is the orientation of the event plane. The latter is reconstructed using the V0 detectors. The calibrated amplitude of the signal measured in each cell of the V0 arrays is used as a weight $w_{\mathrm{cell}}$ in the construction of the flow vector $Q_{2}$ \cite{EventPlaneMethod}

\begin{equation}
Q_{2} = \sum_{j=1}^{N_{\mathrm{cell}}} w_{\mathrm{cell}} \cdot \mathrm{exp} (i 2 \varphi_{\mathrm{cell}})
\label{Eq:Qn}
\end{equation}

where $N_{\mathrm{cell}}$ is the number of cells of the V0 detectors and $\varphi_{\mathrm{cell}}$ is the azimuthal angle of the geometric center of each cell.
In order to account for a non-uniform detector response which can generate a bias in the $\Psi_{2}^{\mathrm{EP}}$ distribution, the components of the $Q_{2}$-vector are adjusted using a re-centering procedure \cite{RecenteringQn} . The orientation of the event plane angle is obtained using the real and imaginary parts of $Q_{2}$  

\begin{equation}
\Psi_{2}^{\mathrm{EP}} = \frac{1}{2} \mathrm{arctan} \left(\frac{Im(Q_{n})}{Re(Q_{n})}    \right)
\label{Eq:Psi2}
\end{equation}

\begin{figure}[hbt]
    \begin{center}
    \includegraphics[width = 0.6\textwidth]{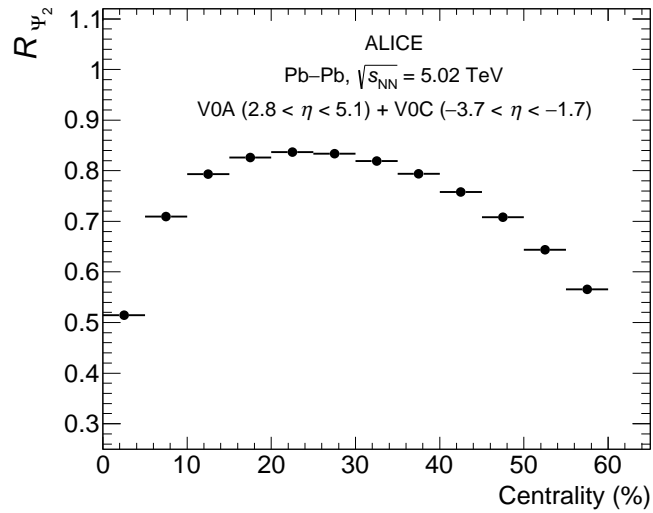}
    \end{center}
    \caption{Event-plane resolution $R_{\Psi_{2}}$ of the second harmonic as a function of the collision centrality.}
    \label{fig:Resolution}
\end{figure}

The event-plane resolution $R_{\Psi_{2}}$ is calculated using the three sub-event correlation technique with charged particles \cite{EventPlaneMethod}

\begin{equation}
R_{\Psi_{2}} = \sqrt{   \frac{ \langle \mathrm{cos} \left( 2 \left( \Psi_{2}^{\mathrm{A}} - \Psi^{\mathrm{B}}_{2}  \right) \right) \rangle \cdot  \langle \mathrm{cos} \left( 2 \left( \Psi_{2}^{\mathrm{A}} - \Psi^{\mathrm{C}}_{2}  \right) \right) \rangle}{ \langle \mathrm{cos} \left( 2 \left( \Psi_{2}^{\mathrm{B}} - \Psi^{\mathrm{C}}_{2}  \right) \right) \rangle}},
\label{Eq:Resolution}
\end{equation}

where A refers to the event plane measured using the V0 detectors, while B and C refer to those obtained in the positive ($\eta>0$) and negative ($\eta<0$) pseudorapidity regions of the TPC. 
For the latter two measurements, a set of reconstructed charged tracks with $0.2\ \leq p_{\mathrm{T}} < 20\ \mathrm{GeV}/\textit{c}$ and $|\eta|<0.8$ is used. Minimal quality criteria are applied to these tracks, such as the requirement of having a number of TPC clusters larger than 70 and a $\chi^{2}/N_{\mathrm{cls}}^{\mathrm{TPC}}<4$. 
The second harmonic event-plane resolution as a function of the collision centrality is shown in Fig.~\ref{fig:Resolution}.

Considering the centrality dependence of $R_{\Psi_{2}}$, the elliptic flow measurements are performed in centrality intervals of 5$\%$ width for the range 0--40$\%$, and of 10$\%$ width for the range 40--60$\%$. The latter two intervals are larger due to the limited number of (anti-)$^{3}$He candidates. 
The resolutions for the centrality ranges 40--50$\%$ and 50--60$\%$ are given by the weighted averages of the resolutions calculated in centrality bins of 5$\%$ width, with the number of charged tracks in the corresponding centrality ranges as a weight.
Finally, the elliptic flow measurements for the wider centrality classes used in this analysis are obtained as weighted averages of the measurements in the smaller centrality ranges

\begin{equation}
v_{2} \left( p_{\mathrm{T}} \right) = \frac{\sum_{i} v_{2}^{i} \left( p_{\mathrm{T}} \right) \cdot N^{i}_{(\mathrm{anti}\mhyphen)^{3}\mathrm{He}}\left( p_{\mathrm{T}} \right)} {\sum_{i} N^{i}_{(\mathrm{anti}\mhyphen)^{3}\mathrm{He}}\left( p_{\mathrm{T}} \right) },
\label{Eq:WeightedAverageEllipticFlow}
\end{equation}

where $v_{2}^{i} \left( p_{\mathrm{T}} \right)$ is the elliptic flow measured in a given \pt range and in the centrality interval $i$, and $N^{i}_{(\mathrm{anti}\mhyphen)^{3}\mathrm{He}}$ is the number of (anti-)$^{3}$He candidates for the same centrality and \pt range.

\section{Systematic uncertainties} 
\label{sec:SystematicUncertainties}

The main sources of systematic uncertainties in this measurement are related to the event selection criteria, track reconstruction, particle identification, occupancy effects in the TPC, and the subtraction of the feed-down contribution from weak decays of hypertritons. Except for the systematic uncertainty due to the event selection, all other contributions are estimated using Monte Carlo (MC) simulations based on the HIJING generator \cite{HIJING}. Simulated events are enriched by an injected sample of (anti-)(hyper-)nuclei generated with a flat \pt distribution in the transverse-momentum range $0 < p_{\mathrm{T}} < 10\ \mathrm{GeV}/\textit{c}$ and a flat rapidity distribution in the range $-1< y < 1$. The interactions of the generated particles with the experimental apparatus are modeled by GEANT 3 \cite{GEANT}. 
The input transverse-momentum distribution of injected (anti-)$^{3}$He is corrected using centrality and \pt-dependent weights to reproduce its measured shape, which is described by the Blast-Wave function. The parameters are taken from the (anti-)$^{3}$He measurement in \PbPb collisions at \snn = 2.76 TeV \cite{3HePbPb} assuming the same spectral shape in \PbPb collisions at \snn = 5.02 TeV.
The systematic uncertainties estimated using the MC simulations are found to be independent on the input parametrization of the (anti-)$^{3}$He spectrum.
A good matching between the distributions of variables used for track selection and particle identification is found between data and MC simulations. This guarantees the reliability of the detector response description and of the systematic uncertainties obtained based on MC simulations.

\subsection{Systematic uncertainties due to the event selection criteria} 
\label{subsec:SystEventSelection}

The effect of different event selection criteria is studied by comparing the $v_{2}$ measurements obtained by varying the selection range of the $z$-coordinate of the primary vertex, using different centrality estimators, selecting events corresponding to opposite magnetic field orientations, using different pile-up rejection criteria, and selecting events with different interaction rates. 
The limited number of (anti-)$^{3}$He candidates prevents the estimation of this source of systematic uncertainties from data since the $v_{2}$ measurements obtained using these different selection criteria are consistent within their statistical uncertainties, i.e. the systematic uncertainties are comparable to or smaller than the statistical ones. The systematic uncertainty related to event selection criteria is assumed to be identical to that of the proton $v_{2}$ measured in \PbPb collisions at \snn = 5.02 TeV and it is taken from Ref. \cite{protonV2}. The total systematic uncertainty due to the event selection is $2.7\%$ and is obtained by adding all contributions in quadrature.

\subsection{Systematic uncertainties due to tracking and particle identification} 
\label{subsec:SystTrackingPID}

The systematic uncertainties due to track reconstruction and particle identification are estimated using MC simulations. This is done to benefit from the larger number of (anti-)$^{3}$He in the simulation as compared to data to reduce the interference between statistical fluctuations and systematic uncertainties.
The same azimuthal asymmetry as measured in data in each centrality and \pt range is artificially created for the injected (anti-)$^{3}$He with respect to a randomly oriented event plane by rejecting a fraction of the out-of-plane (anti-)$^{3}$He. This is done because the injected (anti-)$^{3}$He are produced with $v_{2}=0$ by the MC generator.
The $v_{2}$ of the embedded (anti-)$^{3}$He is then measured using the reconstructed tracks in the simulation. 
Different track selection criteria and signal extraction ranges are used to measure the $v_{2}$, in which the analysis parameters are selected randomly inside a range around the default value using a uniform probability distribution. The different selection criteria are varied simultaneously in order to include the effects of their possible correlations. In each centrality class and for each transverse-momentum range, the measurements obtained using different selection criteria follow a Gaussian distribution whose standard deviation is very similar to the statistical uncertainty, indicating a residual correlation between systematic variations and statistical fluctuations. 
Assuming that the spread of the different measurements is only due to statistical fluctuations, the mean of the Gaussian distribution is considered as the best estimate of the reconstructed $v_{2}$. The difference between the injected $v_{2}$ in the simulation and the mean of the Gaussian spread of the measurements is taken as the systematic uncertainty due to tracking and PID. This uncertainty ranges between 1$\%$ and 4$\%$, depending on \pt and centrality. 
An additional component to the tracking uncertainty originates from the difference between the $v_{2}$ measured using the positive and negative pseudorapidity regions of the TPC. This contribution cannot be estimated from data due to the limited number of (anti-)$^{3}$He and is assumed to be identical to that of the proton $v_{2}$ measurement, which is $2\%$ \cite{protonV2}. The latter is added in quadrature to the systematic uncertainties related to tracking and particle identification.

\subsection{Systematic uncertainty due to occupancy effects in the TPC} 
\label{subsec:SystOccupancy}

Different reconstruction efficiencies for in-plane and out-of-plane particles, due to occupancy effects in the TPC, can create a bias in the $v_{2}$ measurement. This effect is studied using MC simulations by comparing the reconstruction efficiency for different charged-particle multiplicities. The same track selection criteria used in data are applied to the reconstructed tracks in the simulation for the efficiency calculation. The maximum deviation between the reconstruction efficiencies for different multiplicities is $ 0.5\%$, corresponding to a ratio between in-plane and out-of-plane efficiencies of $r=0.995 \pm 0.001$.
The difference between the $v_{2}$ measured assuming $r=1$ and $r=0.995$ corresponds to the maximum variation range of $v_{2}$. The systematic uncertainty from occupancy is then given by this maximum difference divided by $\sqrt{12}$,  assuming a uniform distribution. This uncertainty decreases with increasing \pt and yields at maximum $2\%$ for the centrality range 0--20$\%$ and $0.5\%$ for the centrality ranges 20--40$\%$ and 40--60$\%$.

\subsection{Systematic uncertainty due to the feed-down subtraction} 
\label{subsec:SystFeedDown}

The feed-down systematic uncertainty is due to the unknown $v_{2}$ of (anti-)$^{3}$He from the weak decay of the (anti-)$^3_\Lambda$H. The fraction of secondary (anti-)$^{3}$He from the (anti-)$^3_\Lambda$H decays in the reconstructed track sample is calculated using MC simulations. This fraction is about $6\%$ for the centrality range 0--20$\%$ and $\sim 5\%$ for the centrality ranges 20--40$\%$ and 40--60$\%$, slightly increasing with \pt. 
The relative abundances of (anti-)$^3_\Lambda$H and (anti-)$^{3}$He in the simulation are adjusted to the measured values in \PbPb collisions at \snn = 2.76 TeV \cite{Hypertriton}, assuming that the yield ratio between (anti-)$^3_\Lambda$H and (anti-)$^{3}$He does not differ significantly from that in \PbPb collisions at  \snn = 5.02 TeV, which is not published yet. 
The $v_{2}$ of (anti-)$^{3}$He from the (anti-)$^3_\Lambda$H decay is assumed to be within the range of $\pm 50\%$ with respect to the $v_{2}$ of the inclusive (anti-)$^{3}$He. This variation is selected to provide a conservative estimate of the feed-down uncertainty.
For each of these extremes, the feed-down contribution is subtracted. The systematic uncertainty due to the feed-down subtraction is given by the difference between these two limits divided by $\sqrt{12}$. This uncertainty is $\sim2\%$ in all centrality ranges, almost independent of \pt. 

The different contributions to the systematic uncertainties of this measurement are summarized in \autoref{Table:TotalSystUncertainties}.

\begin{table}[!hbt]
\begin{center}
\centering
\renewcommand{\arraystretch}{1.1}
\caption{Summary of systematic uncertainties. The ranges represent the minimum and maximum uncertainties in the case where the systematic uncertainties depend on \pt and centrality. }
\begin{tabular}{lllc}
\hline
Source of systematic uncertainty &  &   &   value ($\%$)  \\
\hline
     Primary vertex selection     &   &   &   1   \\
     Centrality estimator            &   &    &  1.5    \\
     Magnetic field orientation   &   &   &   1    \\
     Pile-up rejection                  &   &    &  1   \\
     Interaction rate                   &   &    &  1.5   \\
     Tracking and particle identification                 &  &   &    $2-4.5$  \\
     Occupancy in the TPC          &  &   &    $0.5-2$  \\
     Feed-down                           &  &   &    $2$  \\
     \hline
     Total                                   &  &    &   $4-6$    \\
\hline
\end{tabular}
\label{Table:TotalSystUncertainties}
\end{center}
\end{table}

\section{Results} 
\label{sec:Results}

\subsection{Experimental results} 
\label{subsec:ExperimentalResults}

\begin{figure}[!hbt]
    \begin{center}
    \includegraphics[width = 0.6\textwidth]{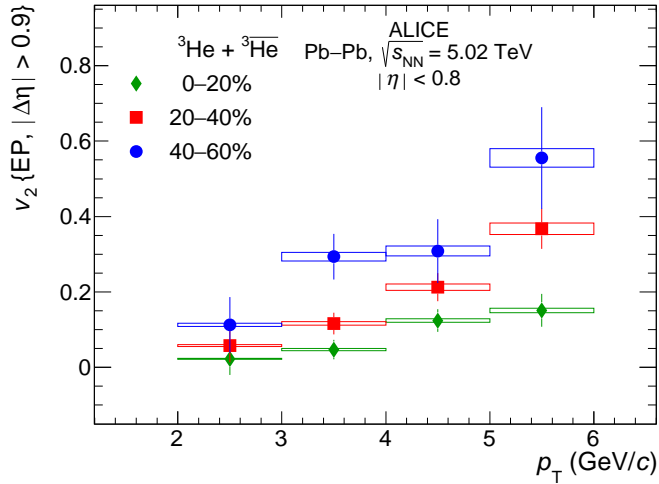}
    \end{center}
    \caption{Elliptic flow ($v_{2}$) of (anti-)$^{3}$He measured in \PbPb collisions at \snn = 5.02 TeV for the centrality classes 0--20$\%$, 20--40$\%$, and 40--60$\%$. The statistical uncertainties are shown as vertical bars, systematic uncertainties as boxes. }
    \label{fig:EllipticFlowHe3}
\end{figure}

The elliptic flow of (anti-)$^{3}$He measured in \PbPb collisions at \snn = 5.02 TeV for the centrality classes 0--20$\%$, 20--40$\%$ and 40--60$\%$ is shown in Fig.~\ref{fig:EllipticFlowHe3} as a function of \pt. The measurement in the transverse-momentum range $2 < p_{\mathrm{T}} < 3\ \mathrm{GeV}/\textit{c}$ is done using only $^{3}\overline{\mathrm{He}}$. An increasing elliptic flow is observed going from central to semi-central collisions, as expected. 
This is due to the increasing azimuthal asymmetry of the overlap region of the colliding nuclei at the initial collision stage, which results in a larger azimuthal asymmetry of the momenta of the final-state particles.
In each centrality class, the elliptic flow increases with \pt in the measured \pt range.

\begin{figure}[!hbt]
    \begin{center}
    \includegraphics[width = 0.98\textwidth]{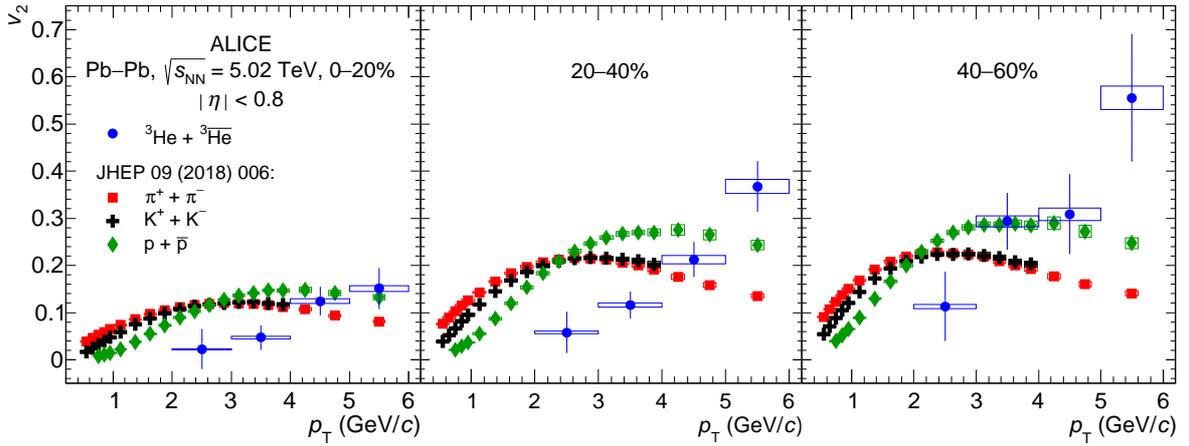}
    \end{center}
    \caption{Comparison between the elliptic flow of (anti-)$^{3}$He measured using the event-plane method and that of pions, kaons, and protons measured using the scalar-product method in \PbPb collisions at \snn = 5.02 TeV for the centrality classes 0--20$\%$ (left), 20--40$\%$ (middle) and 40--60$\%$ (right). See text for details. Vertical bars and boxes represent the statistical and systematic uncertainties, respectively. }
    \label{fig:ComparisonPionsKaonsProtons}
\end{figure}

The (anti-)$^{3}$He elliptic flow is compared to that of pions, kaons, and protons measured using the scalar-product method at the same center-of-mass energy \cite{protonV2} in Fig.~\ref{fig:ComparisonPionsKaonsProtons}. Given the good event-plane resolution shown in Fig.~\ref{fig:Resolution} and the large statistical uncertainties of the (anti-)$^{3}$He $v_{2}$ measurements, the difference between the scalar-product and event-plane method to calculate the (anti-)$^{3}$He elliptic flow is negligible. The $v_{2}$ of pions, kaons, and protons is measured in smaller centrality ranges compared to those used in this analysis. The corresponding $v_{2}$ for the centrality classes 0--20$\%$, 20--40$\%$, and 40--60$\%$ are obtained as weighted averages of the $v_{2}$ measured in smaller centrality classes using the \pt spectra taken from \cite{protonSpectra} as weights. A clear mass ordering is observed for $p_{\mathrm{T}} < 3\ \mathrm{GeV}/\textit{c}$, consistent with the expectations from relativistic hydrodynamics \cite{relativisticHydro}. The $v_{2}$ of (anti-)$^{3}$He shows a slower rise with \pt compared to that of pions, kaons, and protons due to its larger mass.

\begin{figure}[!hbt]
    \begin{center}
    \includegraphics[width = 0.98\textwidth]{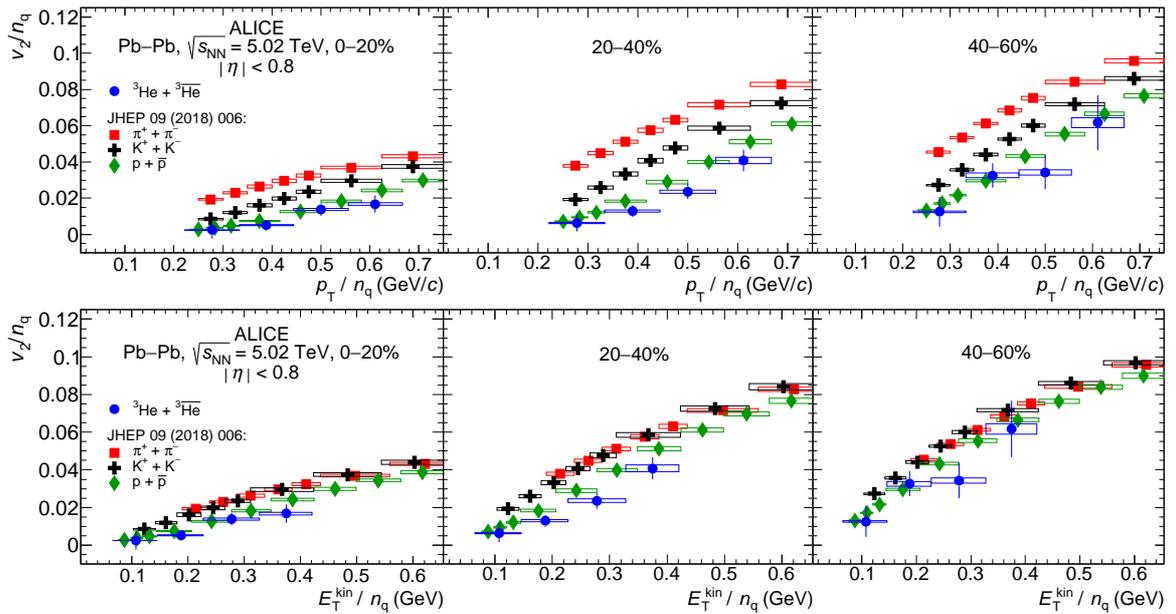}
    \end{center}
    \caption{Comparison between the elliptic flow of pions, kaons, protons, and (anti-)$^{3}$He divided by the number of constituent quarks ($n_{\mathrm{q}}$) as a function of $p_{\mathrm{T}}/n_{\mathrm{q}}$ (upper panels) and transverse kinetic energy per constituent quark $E^{\mathrm{kin}}_{\mathrm{T}}/n_{\mathrm{q}}$ (lower panels) for the centrality classes 0--20$\%$ (left), 20--40$\%$ (middle) and 40--60$\%$ (right). See text for details. Vertical bars and boxes represent the statistical and systematic uncertainties, respectively.}
    \label{fig:nqScaling}
\end{figure}

The comparisons between the measurements of $v_{2}/n_{\mathrm{q}}$ of (anti-)$^{3}$He, pions, kaons, and protons are shown in Fig.~\ref{fig:nqScaling} as a function of $p_{\mathrm{T}}/n_{\mathrm{q}}$ (upper panels), and transverse kinetic energy per constituent quark $E^{\mathrm{kin}}_{\mathrm{T}}/n_{\mathrm{q}}$ (lower panels). 
The transverse kinetic energy is defined as $E^{\mathrm{kin}}_{\mathrm{T}} = \sqrt{m^{2}+p^{2}_{\mathrm{T}}} - m$, where $m$ is the mass of the particle.
The violation of $n_{\mathrm{q}}$ scaling for the measured range of $p_{\mathrm{T}}/n_{\mathrm{q}} \lesssim 0.7\ \mathrm{GeV}/\textit{c}$, already established for the elliptic flow measurements of identified hadrons at the LHC \cite{FlowChargedParticlesLHC,protonV2,deuteronFlowALICE}, is observed also for (anti-)$^{3}$He.
The $n_{\mathrm{q}}$ scaling at larger $p_{\mathrm{T}}/n_{\mathrm{q}}$ cannot be tested with the limited data sample used for this analysis.

\subsection{Model comparisons} 
\label{subsec:ModelComparisons}

The (anti-)$^{3}$He $v_{2}$ measurements are compared with the expectations from the Blast-Wave model and a simple coalescence approach using the same procedure followed in \cite{deuteronFlowALICE}. 

The Blast-Wave predictions are obtained from a simultaneous fit of the $v_{2}$ and the \pt spectra of pions, kaons, and protons measured in \PbPb collisions at \snn = 5.02 TeV \cite{protonV2,protonSpectra} in the transverse-momentum ranges $0.5 \leq p^{\pi}_{\mathrm{T}} < 1\ \mathrm{GeV}/\textit{c}$, $0.7 \leq p^{\mathrm{K}}_{\mathrm{T}} < 2\ \mathrm{GeV}/\textit{c}$, and  $0.7 \leq p^{\mathrm{p}}_{\mathrm{T}} < 2.5\ \mathrm{GeV}/\textit{c}$, respectively, and in the same centrality classes. 
The four parameters of the Blast-Wave fits represent the kinetic freeze-out temperature ($T_{\mathrm{kin}}$), the mean transverse expansion rapidity ($\rho_{0}$), the amplitude of its azimuthal variation ($\rho_{\mathrm{a}}$), and the variation in the azimuthal density of the source ($s_{2}$), as described in \cite{BlastWave2}. The values of the Blast-Wave parameters extracted from the fits are reported in \autoref{Table:BlastWaveParameters} for each centrality interval. The elliptic flow of (anti-)$^{3}$He is calculated using the parameters obtained from the simultaneous fit and the $^{3}$He mass, i.e., assuming the same kinetic freeze-out conditions.

\begin{table}[!hbt]
\begin{center}
\centering
\renewcommand{\arraystretch}{1.1}
\caption{Blast-Wave parameters extracted from the simultaneous fits of the \pt spectra and $v_{2}$ of pions, kaons, and protons. See text for details. }
\begin{tabular}{cccc}
\hline
Fit parameters &     &  Centrality classes  &  \\
  &  0--20$\%$ &  20--40$\%$ &  40--60$\%$ \\
\hline
     $T_{\mathrm{kin}}$ (MeV)             & $106  \pm 1$  & $110  \pm 1$  &  $117  \pm 1$   \\
     $\rho_{0}$ $\times10^{-1}$           & $8.78 \pm 0.01$  & $8.92 \pm 0.02$  &  $7.48 \pm 0.01$   \\
     $\rho_{\mathrm{a}}$ $\times10^{-2}$  & $1.37 \pm 0.01$  & $2.98 \pm 0.01$  &  $3.16 \pm 0.01$   \\
     $s_{2}$  $\times10^{-2}$             & $4.06 \pm 0.01$  & $9.02 \pm 0.01$  &  $1.29 \pm 0.01$   \\
\hline
\end{tabular}
\label{Table:BlastWaveParameters}
\end{center}
\end{table}

The simple coalescence approach used in this context is based on the assumption that the invariant yield of (anti-)$^{3}$He with transverse momentum \pt is proportional to the product of the invariant yields of its constituent nucleons with transverse momentum $p_{\mathrm{T}}/3$ and on isospin symmetry, for which the proton and neutron $v_{2}$ are identical. 
Considering only elliptical anisotropies of the constituent nucleons, i.e. neglecting higher order harmonics, the coalescence predictions are obtained from the elliptic flow of protons $v_{2,\mathrm{p}}$ measured in \PbPb collisions at \snn = 5.02 TeV \cite{protonV2} using the scaling law \cite{ellipticFlowScaling}

\begin{equation}
v_{2,^{3}\mathrm{He}} \left( p_{\mathrm{T}}\right) = \frac{ 3 v_{2,\mathrm{p}} \left( p_{\mathrm{T}}/3\right) + 3 v^{3}_{2,\mathrm{p}} \left( p_{\mathrm{T}}/3\right) }{1 + 6 v^{2}_{2,\mathrm{p}} \left( p_{\mathrm{T}}/3\right)}.
\end{equation}

\begin{figure}[!hbt]
    \begin{center}
    \includegraphics[width = 0.98\textwidth]{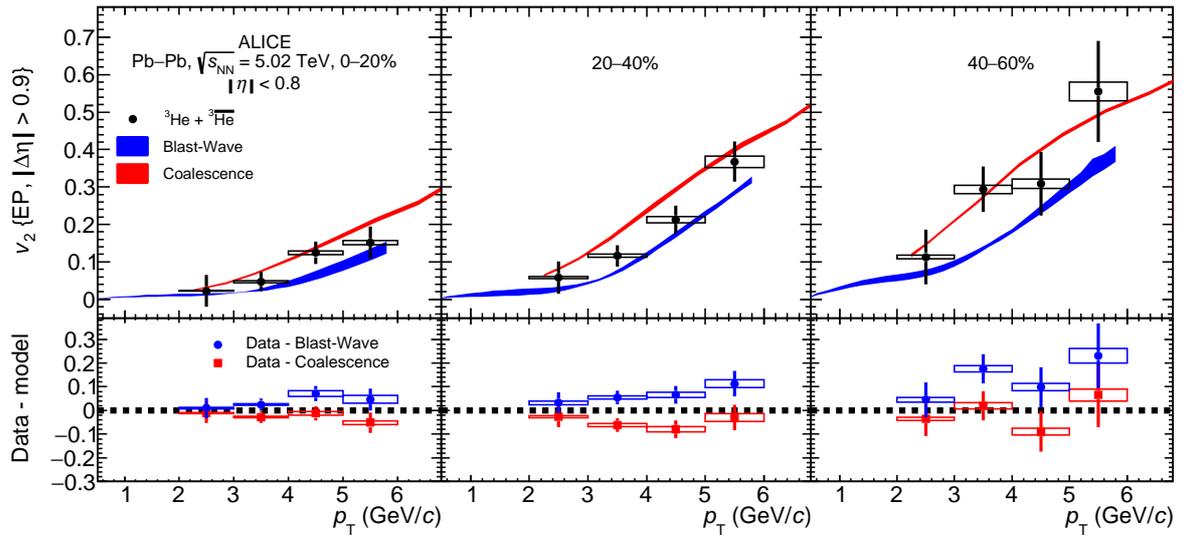}
    \end{center}
    \caption{Elliptic flow of (anti-)$^{3}$He in comparison with the predictions from the Blast-Wave model and a simple coalescence approach for the centrality classes 0--20$\%$ (left), 20--40$\%$ (middle), and 40--60$\%$ (right). The lower panels show the differences between data and models for each centrality range. The statistical uncertainties of the data and the model are added in quadrature. Vertical bars and boxes represent the statistical and systematic uncertainties, respectively.}
    \label{fig:dataVsCoalescenceBW}
\end{figure}

Figure~\ref{fig:dataVsCoalescenceBW} shows the comparison of the (anti-)$^{3}$He $v_{2}$ measurements with the predictions of the Blast-Wave model and the simple coalescence approach. The differences between the data and the model for each centrality interval are shown in the lower panels. These are calculated using the weighted averages of the models in the same \pt intervals of the measurement. For the Blast-Wave model, the \pt spectrum of (anti-)$^{3}$He measured in \PbPb collisions at \snn = 2.76 TeV \cite{3HePbPb} is used as a weight. This is justified considering that the (anti-)$^{3}$He \pt spectrum in \PbPb collisions at \snn = 5.02 TeV is expected to be similar to that at \snn = 2.76 TeV, as observed for lighter hadrons \cite{protonSpectra}. 
The proton spectrum measured in \PbPb collisions at \snn = 5.02 TeV \cite{protonSpectra}, with \pt scaled by $A=3$, is used as a weight for the coalescence model.
The data are located between the two model predictions in all centrality intervals except for more peripheral collisions, where the coalescence expectations are closer to the data.

The Blast-Wave model was found to be consistent with the (anti-)deuteron elliptic flow measured in \PbPb collisions at \snn = 2.76 TeV in the centrality intervals 0--10$\%$, 10--20$\%$ and 20--40$\%$, although the (anti-)deuteron \pt distributions were slightly underestimated for \pt $<$ 2 GeV/$\textit{c}$ in the same centrality intervals \cite{deuteronFlowALICE}. 
Similarly to the results presented in this paper for (anti-)$^{3}$He, the predictions from the simple coalescence model overestimated the (anti-)deuteron $v_{2}$ in all centrality intervals. 
In general, the measurements of (anti-)deuteron and (anti-)$^{3}$He elliptic flow at the LHC consistently indicate that the simple coalescence and Blast-Wave models represent the upper and lower edges of a region where the data are typically located. The (anti-)deuteron elliptic flow measured in \PbPb collisions at \snn = 2.76 TeV is simply closer to the lower side of this region.   

The Blast-Wave model is a simplified parametrization of the system expansion which is typically used to describe the hadron \pt spectra and $v_{2}$ with parameters tuned to data. However, this simple model cannot describe the full collective properties and dynamics of the system. For this, an approach based on relativistic viscous hydrodynamics coupled to an hadronic afterburner is needed. The comparison of the measurement presented in this paper with an actual hydrodynamical simulation is unfortunately not possible because there are no predictions for (anti-)$^{3}$He available.
 
\begin{figure}[!hbt]
    \begin{center}
    \includegraphics[width = 0.6\textwidth]{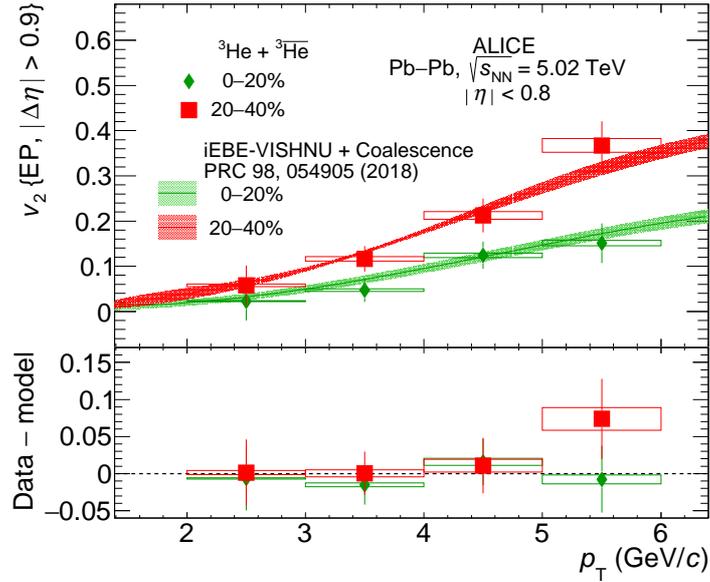}
    \end{center}
    \caption{Elliptic flow of (anti-)$^{3}$He measured in the centrality classes 0--20$\%$ and 20--40$\%$ in comparison with the predictions from a coalescence model based on phase-space distributions of protons and neutrons generated from the iEBE-VISHNU hybrid model with AMPT initial conditions \cite{iEBE_VISHNU}. The model predictions are shown as lines and the bands represent their statistical uncertainties. The differences between data and model are shown in the lower panel for both centrality classes. The statistical uncertainties of the data and the model are added in quadrature. Vertical bars and boxes represent the statistical and systematic uncertainties, respectively.}
    \label{fig:dataVsiEBE_VISHNU}
\end{figure}

The predictions from a more sophisticated coalescence model \cite{iEBE_VISHNU} are compared to the data in the centrality ranges 0--20$\%$ and 20--40$\%$ in Fig.~\ref{fig:dataVsiEBE_VISHNU}. The lower panel shows the differences between the data and the model for these two centrality intervals calculated taking the weighted average of the model in each \pt range, similarly to what is done for the Blast-Wave and the simple coalescence predictions in Fig.~\ref{fig:dataVsCoalescenceBW}. In this model, the coalescence probability is given by the superposition of the wave functions of the coalescing particles, and the Wigner function of the nucleus. The coalescence happens in a flowing medium, i.e., in the rest frame of the fluid cells. This introduces space-momentum correlations absent in the naive coalescence approach. The phase-space distributions of protons and neutrons are generated from the iEBE-VISHNU hybrid model with AMPT initial conditions \cite{iEBE_VISHNU}. 
Although this model underestimates the yield of (anti-)$^{3}$He measured in \PbPb collisions at \snn = 2.76 TeV in the transverse-momentum range of $2\ < p_{\mathrm{T}} < 7\  \mathrm{GeV/\textit{c}}$ by almost a factor of two \cite{iEBE_VISHNU}, it is able to reproduce quantitatively the elliptic flow measurements in the centrality classes 0--20$\%$ and 20--40$\%$ presented here. Moreover, this model provides a good description of the \pt spectra and \pt-differential elliptic flow of protons and deuterons for different centrality intervals in \AuAu collisions at \snn = 200 GeV and in \PbPb collisions at \snn = 2.76 TeV \cite{iEBE_VISHNU}.

\section{Summary} 
\label{sec:Summary}

The first measurement of the (anti-)$^{3}$He elliptic flow in \PbPb collisions at \snn = 5.02 TeV is presented. An increasing trend of $v_{2}$ with \pt and one going from central to semi-central \PbPb collisions is observed. This measurement is compared to that of pions, kaons, and protons at the same center-of-mass energy. A clear mass ordering at low \pt is observed, as expected from relativistic hydrodynamics. The scaling behavior of $v_{2}$ with the number of constituent quarks is violated for the measured range of $p_{\mathrm{T}}/n_{\mathrm{q}} \lesssim 0.7\ \mathrm{GeV}/\textit{c}$ also for (anti-)$^{3}$He, as observed for the $v_{2}$ of lighter particles measured at the LHC. 

The (anti-)$^{3}$He elliptic flow measured in all centrality intervals lies between the predictions from the Blast-Wave model and a simple coalescence approach. This picture is consistent with that of the (anti-)deuteron $v_{2}$ measured in \PbPb collisions at \snn = 2.76 TeV, which was also overestimated by the simple coalescence model, although it was closer to the Blast-Wave predictions. The results on the (anti-)deuteron and (anti-)$^{3}$He elliptic flow measured at the LHC indicate that these two simple models represent upper and lower edges of a region where the elliptic flow of light (anti-)nuclei are typically located.

A more sophisticated coalescence approach based on phase-space distributions of protons and neutrons generated by the iEBE-VISHNU hybrid model with AMPT initial conditions provides a good description of the data in the transverse-momentum interval $2 \leq p_{\mathrm{T}} < 6\ \mathrm{GeV}/\textit{c}$ for the centrality ranges 0--20$\%$ and 20--40$\%$. The same model also provides a good description of the (anti-)deuteron $v_{2}$ measured in \PbPb collisions at \snn = 2.76 TeV. This model, however, fails in the description of the \pt-dependent yield of (anti-)$^{3}$He measured in \PbPb collisions at \snn = 2.76 TeV.

%%%%% acknowledgements - handled by EB chairs 
\newenvironment{acknowledgement}{\relax}{\relax}
\begin{acknowledgement}
\section*{Acknowledgements}
% add specific acknowledgements here 
% ...but please don't remove the line below: funding agencies
% will be acknowledged with a custom tex file handled by EB chairs after Collab Round 2
% Version: 2019-06-11

The ALICE Collaboration would like to thank all its engineers and technicians for their invaluable contributions to the construction of the experiment and the CERN accelerator teams for the outstanding performance of the LHC complex.
The ALICE Collaboration gratefully acknowledges the resources and support provided by all Grid centres and the Worldwide LHC Computing Grid (WLCG) collaboration.
The ALICE Collaboration acknowledges the following funding agencies for their support in building and running the ALICE detector:
A. I. Alikhanyan National Science Laboratory (Yerevan Physics Institute) Foundation (ANSL), State Committee of Science and World Federation of Scientists (WFS), Armenia;
Austrian Academy of Sciences, Austrian Science Fund (FWF): [M 2467-N36] and Nationalstiftung f\"{u}r Forschung, Technologie und Entwicklung, Austria;
Ministry of Communications and High Technologies, National Nuclear Research Center, Azerbaijan;
Conselho Nacional de Desenvolvimento Cient\'{\i}fico e Tecnol\'{o}gico (CNPq), Financiadora de Estudos e Projetos (Finep), Funda\c{c}\~{a}o de Amparo \`{a} Pesquisa do Estado de S\~{a}o Paulo (FAPESP) and Universidade Federal do Rio Grande do Sul (UFRGS), Brazil;
Ministry of Education of China (MOEC) , Ministry of Science \& Technology of China (MSTC) and National Natural Science Foundation of China (NSFC), China;
Ministry of Science and Education and Croatian Science Foundation, Croatia;
Centro de Aplicaciones Tecnol\'{o}gicas y Desarrollo Nuclear (CEADEN), Cubaenerg\'{\i}a, Cuba;
Ministry of Education, Youth and Sports of the Czech Republic, Czech Republic;
The Danish Council for Independent Research | Natural Sciences, the VILLUM FONDEN and Danish National Research Foundation (DNRF), Denmark;
Helsinki Institute of Physics (HIP), Finland;
Commissariat \`{a} l'Energie Atomique (CEA), Institut National de Physique Nucl\'{e}aire et de Physique des Particules (IN2P3) and Centre National de la Recherche Scientifique (CNRS) and R\'{e}gion des  Pays de la Loire, France;
Bundesministerium f\"{u}r Bildung und Forschung (BMBF) and GSI Helmholtzzentrum f\"{u}r Schwerionenforschung GmbH, Germany;
General Secretariat for Research and Technology, Ministry of Education, Research and Religions, Greece;
National Research, Development and Innovation Office, Hungary;
Department of Atomic Energy Government of India (DAE), Department of Science and Technology, Government of India (DST), University Grants Commission, Government of India (UGC) and Council of Scientific and Industrial Research (CSIR), India;
Indonesian Institute of Science, Indonesia;
Centro Fermi - Museo Storico della Fisica e Centro Studi e Ricerche Enrico Fermi and Istituto Nazionale di Fisica Nucleare (INFN), Italy;
Institute for Innovative Science and Technology , Nagasaki Institute of Applied Science (IIST), Japanese Ministry of Education, Culture, Sports, Science and Technology (MEXT) and Japan Society for the Promotion of Science (JSPS) KAKENHI, Japan;
Consejo Nacional de Ciencia (CONACYT) y Tecnolog\'{i}a, through Fondo de Cooperaci\'{o}n Internacional en Ciencia y Tecnolog\'{i}a (FONCICYT) and Direcci\'{o}n General de Asuntos del Personal Academico (DGAPA), Mexico;
Nederlandse Organisatie voor Wetenschappelijk Onderzoek (NWO), Netherlands;
The Research Council of Norway, Norway;
Commission on Science and Technology for Sustainable Development in the South (COMSATS), Pakistan;
Pontificia Universidad Cat\'{o}lica del Per\'{u}, Peru;
Ministry of Science and Higher Education and National Science Centre, Poland;
Korea Institute of Science and Technology Information and National Research Foundation of Korea (NRF), Republic of Korea;
Ministry of Education and Scientific Research, Institute of Atomic Physics and Ministry of Research and Innovation and Institute of Atomic Physics, Romania;
Joint Institute for Nuclear Research (JINR), Ministry of Education and Science of the Russian Federation, National Research Centre Kurchatov Institute, Russian Science Foundation and Russian Foundation for Basic Research, Russia;
Ministry of Education, Science, Research and Sport of the Slovak Republic, Slovakia;
National Research Foundation of South Africa, South Africa;
Swedish Research Council (VR) and Knut \& Alice Wallenberg Foundation (KAW), Sweden;
European Organization for Nuclear Research, Switzerland;
Suranaree University of Technology (SUT), National Science and Technology Development Agency (NSDTA) and Office of the Higher Education Commission under NRU project of Thailand, Thailand;
Turkish Atomic Energy Agency (TAEK), Turkey;
National Academy of  Sciences of Ukraine, Ukraine;
Science and Technology Facilities Council (STFC), United Kingdom;
National Science Foundation of the United States of America (NSF) and United States Department of Energy, Office of Nuclear Physics (DOE NP), United States of America.  
\end{acknowledgement}

%%%%%%%% Bibliography 
\bibliographystyle{utphys}   % Remember we use title in the biblio
\bibliography{bibliography}
%\input {bibliography.tex}  

%%%%%%%%%%%%%%%%%%%%%%%%%%%%%%%%
% Appendices: yours (if any) + authorlist
%%%%%%%%%%%%%%%%%%%%%%%%%%%%%%%%
\newpage
\appendix

%
%\input{} % put your appendices here (if any)
%

%%%%% Authorlist - please do not touch: handled by EB chairs 
\section{The ALICE Collaboration}
\label{app:collab}
% Collaboration: CERN-LHC-ALICE
% Generation Date is 2019-Jun-11

% How to use:
%%%%%%%%% appendix with author list
%\appendix
%\section{The ALICE Collaboration}
%\label{app:collab}
%\input{Alice_Authorslist_XXXX-Axx-XX.tex}
\begingroup
\small
\begin{flushleft}
S.~Acharya\Irefn{org141}\And 
D.~Adamov\'{a}\Irefn{org93}\And 
S.P.~Adhya\Irefn{org141}\And 
A.~Adler\Irefn{org73}\And 
J.~Adolfsson\Irefn{org79}\And 
M.M.~Aggarwal\Irefn{org98}\And 
G.~Aglieri Rinella\Irefn{org34}\And 
M.~Agnello\Irefn{org31}\And 
N.~Agrawal\Irefn{org10}\textsuperscript{,}\Irefn{org48}\textsuperscript{,}\Irefn{org53}\And 
Z.~Ahammed\Irefn{org141}
% ...
\And
\renewcommand\labelenumi{\textsuperscript{\theenumi}~}

\section*{Affiliation notes}
\renewcommand\theenumi{\roman{enumi}}
\begin{Authlist}
\item \Adef{org*}Deceased
\item \Adef{orgI}Dipartimento DET del Politecnico di Torino, Turin, Italy
\item \Adef{orgII}M.V. Lomonosov Moscow State University, D.V. Skobeltsyn Institute of Nuclear, Physics, Moscow, Russia
\item \Adef{orgIII}Department of Applied Physics, Aligarh Muslim University, Aligarh, India
\item \Adef{orgIV}Institute of Theoretical Physics, University of Wroclaw, Poland
\end{Authlist}

\section*{Collaboration Institutes}
\renewcommand\theenumi{\arabic{enumi}~}
\begin{Authlist}
\item \Idef{org1}A.I. Alikhanyan National Science Laboratory (Yerevan Physics Institute) Foundation, Yerevan, Armenia
\item \Idef{org2}Bogolyubov Institute for Theoretical Physics, National Academy of Sciences of Ukraine, Kiev, Ukraine
\item \Idef{org3}Bose Institute, Department of Physics  and Centre for Astroparticle Physics and Space Science (CAPSS), Kolkata, India
\item \Idef{org4}Budker Institute for Nuclear Physics, Novosibirsk, Russia
\item \Idef{org5}California Polytechnic State University, San Luis Obispo, California, United States
\item \Idef{org6}Central China Normal University, Wuhan, China
\item \Idef{org7}Centre de Calcul de l'IN2P3, Villeurbanne, Lyon, France
\item \Idef{org8}Centro de Aplicaciones Tecnol\'{o}gicas y Desarrollo Nuclear (CEADEN), Havana, Cuba
\item \Idef{org9}Centro de Investigaci\'{o}n y de Estudios Avanzados (CINVESTAV), Mexico City and M\'{e}rida, Mexico
\item \Idef{org10}Centro Fermi - Museo Storico della Fisica e Centro Studi e Ricerche ``Enrico Fermi', Rome, Italy
\item \Idef{org11}Chicago State University, Chicago, Illinois, United States
\item \Idef{org12}China Institute of Atomic Energy, Beijing, China
\item \Idef{org13}Chonbuk National University, Jeonju, Republic of Korea
\item \Idef{org14}Comenius University Bratislava, Faculty of Mathematics, Physics and Informatics, Bratislava, Slovakia
\item \Idef{org15}COMSATS University Islamabad, Islamabad, Pakistan
\item \Idef{org16}Creighton University, Omaha, Nebraska, United States
\item \Idef{org17}Department of Physics, Aligarh Muslim University, Aligarh, India
\item \Idef{org18}Department of Physics, Pusan National University, Pusan, Republic of Korea
\item \Idef{org19}Department of Physics, Sejong University, Seoul, Republic of Korea
\item \Idef{org20}Department of Physics, University of California, Berkeley, California, United States
\item \Idef{org21}Department of Physics, University of Oslo, Oslo, Norway
\item \Idef{org22}Department of Physics and Technology, University of Bergen, Bergen, Norway
\item \Idef{org23}Dipartimento di Fisica dell'Universit\`{a} 'La Sapienza' and Sezione INFN, Rome, Italy
\item \Idef{org24}Dipartimento di Fisica dell'Universit\`{a} and Sezione INFN, Cagliari, Italy
\item \Idef{org25}Dipartimento di Fisica dell'Universit\`{a} and Sezione INFN, Trieste, Italy
\item \Idef{org26}Dipartimento di Fisica dell'Universit\`{a} and Sezione INFN, Turin, Italy
\item \Idef{org27}Dipartimento di Fisica e Astronomia dell'Universit\`{a} and Sezione INFN, Bologna, Italy
\item \Idef{org28}Dipartimento di Fisica e Astronomia dell'Universit\`{a} and Sezione INFN, Catania, Italy
\item \Idef{org29}Dipartimento di Fisica e Astronomia dell'Universit\`{a} and Sezione INFN, Padova, Italy
\item \Idef{org30}Dipartimento di Fisica `E.R.~Caianiello' dell'Universit\`{a} and Gruppo Collegato INFN, Salerno, Italy
\item \Idef{org31}Dipartimento DISAT del Politecnico and Sezione INFN, Turin, Italy
\item \Idef{org32}Dipartimento di Scienze e Innovazione Tecnologica dell'Universit\`{a} del Piemonte Orientale and INFN Sezione di Torino, Alessandria, Italy
\item \Idef{org33}Dipartimento Interateneo di Fisica `M.~Merlin' and Sezione INFN, Bari, Italy
\item \Idef{org34}European Organization for Nuclear Research (CERN), Geneva, Switzerland
\item \Idef{org35}Faculty of Electrical Engineering, Mechanical Engineering and Naval Architecture, University of Split, Split, Croatia
\item \Idef{org36}Faculty of Engineering and Science, Western Norway University of Applied Sciences, Bergen, Norway
\item \Idef{org37}Faculty of Nuclear Sciences and Physical Engineering, Czech Technical University in Prague, Prague, Czech Republic
\item \Idef{org38}Faculty of Science, P.J.~\v{S}af\'{a}rik University, Ko\v{s}ice, Slovakia
\item \Idef{org39}Frankfurt Institute for Advanced Studies, Johann Wolfgang Goethe-Universit\"{a}t Frankfurt, Frankfurt, Germany
\item \Idef{org40}Gangneung-Wonju National University, Gangneung, Republic of Korea
\item \Idef{org41}Gauhati University, Department of Physics, Guwahati, India
\item \Idef{org42}Helmholtz-Institut f\"{u}r Strahlen- und Kernphysik, Rheinische Friedrich-Wilhelms-Universit\"{a}t Bonn, Bonn, Germany
\item \Idef{org43}Helsinki Institute of Physics (HIP), Helsinki, Finland
\item \Idef{org44}High Energy Physics Group,  Universidad Aut\'{o}noma de Puebla, Puebla, Mexico
\item \Idef{org45}Hiroshima University, Hiroshima, Japan
\item \Idef{org46}Hochschule Worms, Zentrum  f\"{u}r Technologietransfer und Telekommunikation (ZTT), Worms, Germany
\item \Idef{org47}Horia Hulubei National Institute of Physics and Nuclear Engineering, Bucharest, Romania
\item \Idef{org48}Indian Institute of Technology Bombay (IIT), Mumbai, India
\item \Idef{org49}Indian Institute of Technology Indore, Indore, India
\item \Idef{org50}Indonesian Institute of Sciences, Jakarta, Indonesia
\item \Idef{org51}INFN, Laboratori Nazionali di Frascati, Frascati, Italy
\item \Idef{org52}INFN, Sezione di Bari, Bari, Italy
\item \Idef{org53}INFN, Sezione di Bologna, Bologna, Italy
\item \Idef{org54}INFN, Sezione di Cagliari, Cagliari, Italy
\item \Idef{org55}INFN, Sezione di Catania, Catania, Italy
\item \Idef{org56}INFN, Sezione di Padova, Padova, Italy
\item \Idef{org57}INFN, Sezione di Roma, Rome, Italy
\item \Idef{org58}INFN, Sezione di Torino, Turin, Italy
\item \Idef{org59}INFN, Sezione di Trieste, Trieste, Italy
\item \Idef{org60}Inha University, Incheon, Republic of Korea
\item \Idef{org61}Institut de Physique Nucl\'{e}aire d'Orsay (IPNO), Institut National de Physique Nucl\'{e}aire et de Physique des Particules (IN2P3/CNRS), Universit\'{e} de Paris-Sud, Universit\'{e} Paris-Saclay, Orsay, France
\item \Idef{org62}Institute for Nuclear Research, Academy of Sciences, Moscow, Russia
\item \Idef{org63}Institute for Subatomic Physics, Utrecht University/Nikhef, Utrecht, Netherlands
\item \Idef{org64}Institute of Experimental Physics, Slovak Academy of Sciences, Ko\v{s}ice, Slovakia
\item \Idef{org65}Institute of Physics, Homi Bhabha National Institute, Bhubaneswar, India
\item \Idef{org66}Institute of Physics of the Czech Academy of Sciences, Prague, Czech Republic
\item \Idef{org67}Institute of Space Science (ISS), Bucharest, Romania
\item \Idef{org68}Institut f\"{u}r Kernphysik, Johann Wolfgang Goethe-Universit\"{a}t Frankfurt, Frankfurt, Germany
\item \Idef{org69}Instituto de Ciencias Nucleares, Universidad Nacional Aut\'{o}noma de M\'{e}xico, Mexico City, Mexico
\item \Idef{org70}Instituto de F\'{i}sica, Universidade Federal do Rio Grande do Sul (UFRGS), Porto Alegre, Brazil
\item \Idef{org71}Instituto de F\'{\i}sica, Universidad Nacional Aut\'{o}noma de M\'{e}xico, Mexico City, Mexico
\item \Idef{org72}iThemba LABS, National Research Foundation, Somerset West, South Africa
\item \Idef{org73}Johann-Wolfgang-Goethe Universit\"{a}t Frankfurt Institut f\"{u}r Informatik, Fachbereich Informatik und Mathematik, Frankfurt, Germany
\item \Idef{org74}Joint Institute for Nuclear Research (JINR), Dubna, Russia
\item \Idef{org75}Korea Institute of Science and Technology Information, Daejeon, Republic of Korea
\item \Idef{org76}KTO Karatay University, Konya, Turkey
\item \Idef{org77}Laboratoire de Physique Subatomique et de Cosmologie, Universit\'{e} Grenoble-Alpes, CNRS-IN2P3, Grenoble, France
\item \Idef{org78}Lawrence Berkeley National Laboratory, Berkeley, California, United States
\item \Idef{org79}Lund University Department of Physics, Division of Particle Physics, Lund, Sweden
\item \Idef{org80}Nagasaki Institute of Applied Science, Nagasaki, Japan
\item \Idef{org81}Nara Women{'}s University (NWU), Nara, Japan
\item \Idef{org82}National and Kapodistrian University of Athens, School of Science, Department of Physics , Athens, Greece
\item \Idef{org83}National Centre for Nuclear Research, Warsaw, Poland
\item \Idef{org84}National Institute of Science Education and Research, Homi Bhabha National Institute, Jatni, India
\item \Idef{org85}National Nuclear Research Center, Baku, Azerbaijan
\item \Idef{org86}National Research Centre Kurchatov Institute, Moscow, Russia
\item \Idef{org87}Niels Bohr Institute, University of Copenhagen, Copenhagen, Denmark
\item \Idef{org88}Nikhef, National institute for subatomic physics, Amsterdam, Netherlands
\item \Idef{org89}NRC Kurchatov Institute IHEP, Protvino, Russia
\item \Idef{org90}NRC «Kurchatov Institute»  - ITEP, Moscow, Russia
\item \Idef{org91}NRNU Moscow Engineering Physics Institute, Moscow, Russia
\item \Idef{org92}Nuclear Physics Group, STFC Daresbury Laboratory, Daresbury, United Kingdom
\item \Idef{org93}Nuclear Physics Institute of the Czech Academy of Sciences, \v{R}e\v{z} u Prahy, Czech Republic
\item \Idef{org94}Oak Ridge National Laboratory, Oak Ridge, Tennessee, United States
\item \Idef{org95}Ohio State University, Columbus, Ohio, United States
\item \Idef{org96}Petersburg Nuclear Physics Institute, Gatchina, Russia
\item \Idef{org97}Physics department, Faculty of science, University of Zagreb, Zagreb, Croatia
\item \Idef{org98}Physics Department, Panjab University, Chandigarh, India
\item \Idef{org99}Physics Department, University of Jammu, Jammu, India
\item \Idef{org100}Physics Department, University of Rajasthan, Jaipur, India
\item \Idef{org101}Physikalisches Institut, Eberhard-Karls-Universit\"{a}t T\"{u}bingen, T\"{u}bingen, Germany
\item \Idef{org102}Physikalisches Institut, Ruprecht-Karls-Universit\"{a}t Heidelberg, Heidelberg, Germany
\item \Idef{org103}Physik Department, Technische Universit\"{a}t M\"{u}nchen, Munich, Germany
\item \Idef{org104}Politecnico di Bari, Bari, Italy
\item \Idef{org105}Research Division and ExtreMe Matter Institute EMMI, GSI Helmholtzzentrum f\"ur Schwerionenforschung GmbH, Darmstadt, Germany
\item \Idef{org106}Rudjer Bo\v{s}kovi\'{c} Institute, Zagreb, Croatia
\item \Idef{org107}Russian Federal Nuclear Center (VNIIEF), Sarov, Russia
\item \Idef{org108}Saha Institute of Nuclear Physics, Homi Bhabha National Institute, Kolkata, India
\item \Idef{org109}School of Physics and Astronomy, University of Birmingham, Birmingham, United Kingdom
\item \Idef{org110}Secci\'{o}n F\'{\i}sica, Departamento de Ciencias, Pontificia Universidad Cat\'{o}lica del Per\'{u}, Lima, Peru
\item \Idef{org111}Shanghai Institute of Applied Physics, Shanghai, China
\item \Idef{org112}St. Petersburg State University, St. Petersburg, Russia
\item \Idef{org113}Stefan Meyer Institut f\"{u}r Subatomare Physik (SMI), Vienna, Austria
\item \Idef{org114}SUBATECH, IMT Atlantique, Universit\'{e} de Nantes, CNRS-IN2P3, Nantes, France
\item \Idef{org115}Suranaree University of Technology, Nakhon Ratchasima, Thailand
\item \Idef{org116}Technical University of Ko\v{s}ice, Ko\v{s}ice, Slovakia
\item \Idef{org117}Technische Universit\"{a}t M\"{u}nchen, Excellence Cluster 'Universe', Munich, Germany
\item \Idef{org118}The Henryk Niewodniczanski Institute of Nuclear Physics, Polish Academy of Sciences, Cracow, Poland
\item \Idef{org119}The University of Texas at Austin, Austin, Texas, United States
\item \Idef{org120}Universidad Aut\'{o}noma de Sinaloa, Culiac\'{a}n, Mexico
\item \Idef{org121}Universidade de S\~{a}o Paulo (USP), S\~{a}o Paulo, Brazil
\item \Idef{org122}Universidade Estadual de Campinas (UNICAMP), Campinas, Brazil
\item \Idef{org123}Universidade Federal do ABC, Santo Andre, Brazil
\item \Idef{org124}University of Cape Town, Cape Town, South Africa
\item \Idef{org125}University of Houston, Houston, Texas, United States
\item \Idef{org126}University of Jyv\"{a}skyl\"{a}, Jyv\"{a}skyl\"{a}, Finland
\item \Idef{org127}University of Liverpool, Liverpool, United Kingdom
\item \Idef{org128}University of Science and Technology of China, Hefei, China
\item \Idef{org129}University of South-Eastern Norway, Tonsberg, Norway
\item \Idef{org130}University of Tennessee, Knoxville, Tennessee, United States
\item \Idef{org131}University of the Witwatersrand, Johannesburg, South Africa
\item \Idef{org132}University of Tokyo, Tokyo, Japan
\item \Idef{org133}University of Tsukuba, Tsukuba, Japan
\item \Idef{org134}Universit\'{e} Clermont Auvergne, CNRS/IN2P3, LPC, Clermont-Ferrand, France
\item \Idef{org135}Universit\'{e} de Lyon, Universit\'{e} Lyon 1, CNRS/IN2P3, IPN-Lyon, Villeurbanne, Lyon, France
\item \Idef{org136}Universit\'{e} de Strasbourg, CNRS, IPHC UMR 7178, F-67000 Strasbourg, France, Strasbourg, France
\item \Idef{org137}Universit\'{e} Paris-Saclay Centre d'Etudes de Saclay (CEA), IRFU, D\'{e}partment de Physique Nucl\'{e}aire (DPhN), Saclay, France
\item \Idef{org138}Universit\`{a} degli Studi di Foggia, Foggia, Italy
\item \Idef{org139}Universit\`{a} degli Studi di Pavia, Pavia, Italy
\item \Idef{org140}Universit\`{a} di Brescia, Brescia, Italy
\item \Idef{org141}Variable Energy Cyclotron Centre, Homi Bhabha National Institute, Kolkata, India
\item \Idef{org142}Warsaw University of Technology, Warsaw, Poland
\item \Idef{org143}Wayne State University, Detroit, Michigan, United States
\item \Idef{org144}Westf\"{a}lische Wilhelms-Universit\"{a}t M\"{u}nster, Institut f\"{u}r Kernphysik, M\"{u}nster, Germany
\item \Idef{org145}Wigner Research Centre for Physics, Hungarian Academy of Sciences, Budapest, Hungary
\item \Idef{org146}Yale University, New Haven, Connecticut, United States
\item \Idef{org147}Yonsei University, Seoul, Republic of Korea
\end{Authlist}
\endgroup
  
\end{document}